# Ab Initio Calculations
# of the Walls Shear Strength of Carbon Nanotubes

**E. Bichoutskaia[a], O. V. Ershova[b], Yu. E. Lozovik[c], and A. M. Popov[c]***
[a]Department of Chemistry, University of Nottingham, Nottingham NG7 2RD, United Kingdom
[b]Moscow Institute of Physics and Technology, Dolgoprudnyi, Moscow oblast, 141701 Russia
[c]Institute of Spectroscopy, Russian Academy of Sciences, Troitsk, Moscow oblast, 142190 Russia
*e-mail: am-popov@isan.troitsk.ru

**Abstract**—The dependence of the energy of interwall interaction in double-walled carbon nanotubes (DWNT) on the relative position of walls has been calculated using the density functional method. This dependence is used to evaluate forces that are necessary for the relative telescopic motion of walls and to calculate the shear strength of DWNT for the relative sliding of walls along the nanotube axis and for their relative rotation about this axis. The possibility of experimental verification of the obtained results is discussed.

The possibility of the relative motion of walls (shells) in multi-walled carbon nanotubes (MWNT) [1] shows good prospects for their use as mobile elements of nanoelectromechanical systems (NEMS). Recently, nanomotors with shafts and sleeves representing different walls of MWNT [2] and a memory cell switch based on the relative motion of walls along the MWNT axis [3] were obtained in experiments. A number of other NEMS based on the relative motion of walls in MWNT were also proposed, including a gigahertz oscillator, Brownian nanomotor, and bolt–nut couples. The operation of NEMS playing the role of a variable nanoresistor, strain nanosensor, and electromechanical nanothermometer is based on the dependence of the conductivity and interwall interaction energy on the relative position of walls. The schemes and operation principles of these NEMS were recently reviewed in [4]. Thus, investigations of the relative motion of walls in MWNT are of considerable importance for the creation and development of NEMS.

However, now there is a lack of reliable and detailed experimental data not only on the interaction between walls of MWNT, but even on the interaction between graphite layers. For example, both the available experimental values and the results of theoretical calculations for the interaction of graphite layers exhibit scatter within two orders of magnitude (see, e.g., [5] and references therein). Only a few experiments were devoted to measurements of the shear strength of nanotubes in the axial direction and only the upper limit of this value (~0.04 MPa) was estimated [1].

As for theoretical investigations, the potential barriers for the relative motion of walls were calculated for a large number of double-walled carbon nanotubes (DWNTs) using semiempirical potentials [6–8]. For some DWNT, the calculations were performed using nonempirical (ab initio) methods [9–13]. Table I summarizes the values of barriers for the relative motion of walls in (5, 5)@(10, 10) DWNT, which were calculated using various methods.

The only available experimental data that can currently be used to verify the adequacy of ab initio calculations and fit the parameters of semiempirical calculations of the interwall interaction in DWNT are the structural parameters, energy characteristics, and elastic properties of graphite. Some semiempirical [6,8] and ab initio [12,13] methods used in the calculations of barriers for the relative motion of walls in (5, 5)@(10, 10) DWNT give the values of graphite characteristics that are in good agreement with experiment. Nevertheless, the published values of these barriers (see Table 1) still show that the results obtained by various methods differ by orders of magnitude. Thus, the further development of theoretical methods for the investigation of the interaction between walls of MWNT requires experimental determination of the values of barriers for the relative motion of walls. Then, a comparison of the experimentally measured and calculated values of these barriers can be used as a criterion for the adequacy of calculation methods.

In this Letter, the results of ab initio calculations of the barriers are used to evaluate the wall shear strength of DWNT. The possibility of experimental verification of the obtained results is discussed.
According to the results of ab initio calculations for (5,5)@(10,10) DWNT [13] and semiempirical calculations [8], the energy $U$ of the interaction of commensurate nonchiral walls of DWNT (($n1,n1$)@($n2,n2$) and ($n1,0$)@($n2,0$)) as a function of their relative position can be interpolated

**Table 1** Values of potential barriers $\Delta U_z$ (for the relative axial sliding of walls) and $\Delta U_\phi$ (for the relative rotation around the axis) per atom of outer wall in (5, 5)@(10, 10) DWNT and the ratio $\gamma_b = \Delta U_\phi / \Delta U_z$

| Ref. | $\Delta U_z$, meV/atom | $\Delta U_\phi$, meV/atom | $\gamma_b$ |
|---|---|---|---|
| [6][a] | 0.008 | 0.025 | 3.13 |
| [8][b] | 0.00745 | 0.0144 | 2.90 |
| [7][c] | 7.5 | 8.7 | 1.16 |
| [11][d] | 0.128 | 0.438 | 3.47 |
| [10][d] | – | 1.2 | – |
| [9][e] | 0.35 | 0.78 | 2.26 |
| [12][f] | 0.125 | 0.259 | 2.08 |
| [13][g] | 0.138 | 0.407 | 2.85 |

**Note**: [a] Lennard_Jones potential in optimized wall structure; [b] Lennard_Jones potential in nonoptimized wall structure; [c] Crespi–Kolmogorov potential in optimized wall structure; [d] tight binding model; [e] density functional method in local density approximation (plane wave basis set); [f] density functional method in local density approximation (*pdpp*-basis set); [g] density functional method in local density approximation (*dddd*-basis set).

using the first two harmonic terms of expansion into the Fourier series

$$U(\phi, z) = U_0 - \frac{\Delta U_\phi}{2} \cos\left(\frac{2\pi}{\delta_\phi}\phi\right) - \frac{\Delta U_z}{2} \cos\left(\frac{2\pi}{\delta_z}z\right), \tag{1}$$

where $\varphi$ is the angle of the relative rotation of walls about the nanotube axis, z is their relative sliding along the axis, $U_0$ is the average energy of the interwall interaction, and $\Delta U_\phi$ and $\Delta U_z$ are the potential barriers for the relative rotation and sliding of walls, respectively. For DWNT with commensurate walls, the values of $U_0$, $\Delta U_\phi$, and $\Delta U_z$ are proportional to the length of the walls' overlap. In this work $U_0$ $\Delta U_z$, and $\Delta U_\phi$ are calculated for infinite DWNT and normalized per one atom of outer wall. The periods of rotation and sliding between equivalent positions are defined as a $\delta_\phi = \pi N / n_1 n_2$ and $\delta_z = l_c / 2$, where $N$ is the greatest common divisor of $n_1$ and $n_2$ and $l_c$ is the nanotube unit cell length. For DWNT with walls possessing common elements of rotational symmetry ($N=n_1$), the potential barrier $\Delta U_\phi$ for the relative rotation can be quite large. This is illustrated in Table I for (5,5)@(10,10) DWNT with common elements of walls' rotational symmetry ($N=5$). In contrast, for DWNT without common elements of walls' rotational symmetry, the dependence of the interwall interaction energy on $\varphi$ is very weak and the second term in expansion (1) can be ignored (see also review [4]).

In order to check for the adequacy of expansion (1), we have calculated ab initio the interwall interaction energy for (6,6)@(11,11) and (9, 0)@(18, 0) DWNT at a fixed rotation angle $\varphi$ and five values of the relative axial displacement z, including the global extrema and saddle points. The calculations were performed using the density functional method in the local density approximation as implemented in AIMPRO software [14]. This approach evaluates the energy of interaction between graphite layers at 35 meV/atom [15] (in agreement with the experimental value of 35±10 meV/atom [5]) and reproduces the elastic and electronic properties of graphite sensitive to the interlayer interaction. The Brillouin zone was described using 18 and 15 *k*-points in the principal axis direction for the (6,6)@(11,11) and (9,0)@(18,0) DWNT, respectively. Details of the calculation procedure are described elsewhere [13].

Figures 1a and 1b present the calculated dependences of the interwall interaction energy on the relative displacement of walls along the nanotube axis for (6,6)@(11,11) and (9,0)@(18,0) DWNT, respectively. These plots are well interpolated by cosine functions, thus demonstrating the adequacy of expansion (1). Therefore, we can use this expansion for calculating the characteristics of a series of $(n_1, n_1)@(n_2, n_2)$ and $(n_1, 0)@(n_2, 0)$ DWNT. Using the aforementioned interpolation, we estimated the potential barriers for the relative sliding of walls along the nanotube axis as $\Delta U_z = 0.19 \pm 0.01$ meV/atom for (6,6)@(11,11) DWNT and $\Delta U_z = 1.71 \pm 0.04$ meV/atom for (9,0)@(18,0) DWNT. For the other DWNT, the energy parameters ($U_0$, $\Delta U_z$, and $\Delta U_\phi$) and structural data for subsequent calculations were taken from [13].

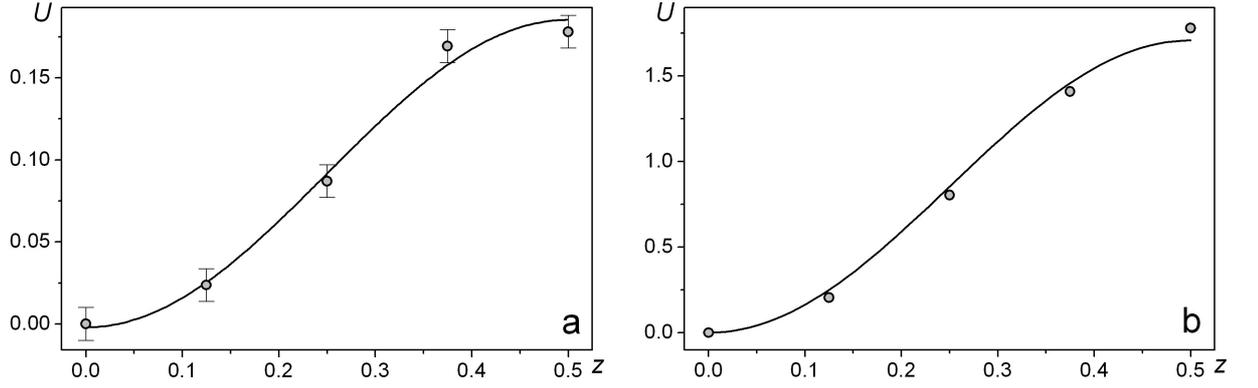

**Fig. 1.** Plots of the interwall interaction energy $U$ (measured from the minimum) versus relative axial displacement of walls $z$ (expressed in units of the sliding period $\delta_z$) for (a) (6,6)@(11,11) and (b) (9,0)@(18,0) DWNT. The points present the results of ab initio calculations, solid curves show the interpolation by cosine functions.

The force of interaction between DWNT walls can be separated into two components, including a capillary force ($F_c$), which arises during a telescopic protrusion of the inner wall (and tends to pull this wall back), and a static friction force related to a particular relief of the $U(\varphi,z)$ potential surface. The average value of the capillary force is given by the following expression:

$$\langle F_c \rangle = \left\langle \frac{dU}{dL_{ov}} \right\rangle = \frac{U_0 4n_2}{l_c}, \qquad (2)$$

where $L_{ov}$ is the walls' overlap length and $4n_2$ is the number of atoms per unit cell in the outer wall. The values of $\langle F_c \rangle$ calculated for various DWNT are presented in Table 2.

The maximum values of the static friction force $F_z$ and $F_\varphi$ for the relative sliding and rotation of walls, respectively, are determined using expansion (1). These values are defined as follows [13]:

$$F_z = \frac{4n_2 \pi L_{ov} \Delta U_z}{\delta_z l_c}, \qquad F_\phi = \frac{4n_2 \pi L_{ov} \Delta U_z}{\delta_\phi R_m l_c}, \qquad (3)$$

where $R_m$ is the radius of the moving wall.

**Table 2.** Calculated values of the average capillary force $\langle F_c \rangle$ and the shear strengths for the relative sliding of walls along the nanotube axis ($M_z$) and for their relative rotation about this axis ($M_\phi$) in various DWNT

| DWNT | $\langle F_c \rangle$, nN | $M_z$, MPa | $M_\phi$, MPa |
| --- | --- | --- | --- |
| (4,4)@(10,10) | 0.3484 | $4.0 \pm 1.7$ | |
| (5,5)@(11,11) | 0.3856 | $5.3 \pm 1.5$ | |
| (6,6)@(12,12) | 0.4305 | $6.5 \pm 1.3$ | $2.4 \pm 1.3$ |
| (5,5)@(10,10) | 0.6253 | $30.4 \pm 1.6$ | $102.2 \pm 1.6$ |
| (6,6)@(11,11) | 0.6954 | $38.4 \pm 2.3$ | |
| (7,7)@(12,12) | 0.7747 | $44.2 \pm 1.2$ | |
| (9,0)@(18,0) | 0.6952 | $215.1 \pm 0.5$ | $12.1 \pm 1.5$ |
| (10,0)@(20,0) | 0.5078 | $66.5 \pm 0.4$ | $11.7 \pm 1.4$ |

The shear strengths of DWNT for the relative sliding of walls along the nanotube axis and for their relative rotation about this axis are defined as follows:

$$M_z = \frac{F_z}{S}, \qquad M_\phi = \frac{F_\phi}{S}, \qquad (4)$$

where

$$S = 2\pi L_{ov}\left(\frac{R_1 + R_2}{2}\right) = \pi L_{ov}(R_1 + R_2), \qquad (5)$$

is the walls' overlap area and $R_1$ and $R_2$ are the radii of the inner and outer walls, respectively. The values of shear strengths calculated for various DWNT are presented in Table 2.

Now let us discuss the possibility of experimental verification of the above results. By now, the upper boundary of the shear strength for the relative axial sliding of nanotube walls was evaluated by atomic force microscopy (AFM) as $M_z < 0.04$ MPa [1]. In most MWNT with both commensurate and incommensurate chiral walls, the barriers for the relative axial sliding of adjacent walls (and hence the corresponding shear strengths) are negligibly small (see, e.g., review [4]). Only the DWNT with nonchiral commensurate walls considered in this study possess significant barriers for the relative sliding of adjacent layers [4,7,8,13]. In the experiment [1], the shear strength was determined only for one pair of adjacent walls in an MWNT. Moreover, this was a pair that possesses the smallest value of this shear strength as compared to the other pairs. Unfortunately, the chirality indices of interacting walls in MWNT studied in [1] were not determined.

Thus, we believe that the available experimental estimate of the upper boundary of shear strengths for the relative axial sliding of walls refers to the incommensurate or commensurate chiral walls. The values of shear strengths obtained in our calculations are several orders of magnitude greater than the upper boundary (0.04 MPa) provided by the AFM for the relative sliding of adjacent walls with undetermined chirality indices. The chirality indices of both layers in DWNT can be determined with the aid of electron diffractometry [16]. Therefore, the capillary forces and shear strengths calculated in this study can be simultaneously determined by AFM measurements of the interwall interaction force as a function of the length of telescopic protrusion of the inner wall in DWNT with preliminarily determined chirality indices. These measurements are necessary both for progress in calculations of the interwall interactions in MWNT and for the development of NEMS based on these interactions.

**Acknowledgments.** This study was supported by the Russian Foundation for Basic Research, project nos. 08-02-90049-Bel and 08-2-00685.